\documentclass[secnumarabic, twocolumn, graphics,floatfix, superscriptaddress,tightenlines, aps, prb]{revtex4-1}
\usepackage[dvips]{graphicx}
\usepackage{dcolumn}
\usepackage{bm}
\usepackage{color}

\begin{document}

\title{Nuclear spin warm-up in bulk  n-GaAs}

\author{M.~Kotur}
\affiliation{Ioffe Physico-Technical Institute of the RAS, 194021 St.Petersburg, Russia}

\author{R. I. Dzhioev}
\affiliation{Ioffe Physico-Technical Institute of the RAS, 194021 St.Petersburg, Russia}

\author{M.~Vladimirova}
\affiliation{Laboratoire Charles Coulomb, UMR 5221 CNRS/ Universit\'{e}  de Montpellier,
F-34095, Montpellier, France}

\author{B.~Jouault}
\affiliation{Laboratoire Charles Coulomb, UMR 5221 CNRS/ Universit\'{e}  de Montpellier,
F-34095, Montpellier, France}

\author{V. L. Korenev}
\affiliation{Ioffe Physico-Technical Institute of the RAS, 194021 St.Petersburg, Russia}

\author{K.~V.~Kavokin}
\affiliation{Spin Optics Laboratory, St-Petersburg State
University, 1, Ulianovskaya, St-Peterbsurg, 198504, Russia}
\affiliation{Ioffe Physico-Technical Institute of the RAS, 194021 St.Petersburg, Russia}

\begin{abstract}
We show that the spin-lattice relaxation in n-type insulating GaAs  is dramatically accelerated at low
magnetic fields.
The origin of this effect, that cannot be explained in terms of well-known diffusion-limited hyperfine relaxation, 
is found in the quadrupole relaxation, induced by fluctuating donor charges.
Therefore, quadrupole relaxation, that governs low field nuclear spin relaxation in semiconductor quantum dots,
but was so far supposed to be harmless to bulk nuclei spins in the absence of optical pumping can be studied and harnessed in much simpler model environment of n-GaAs bulk crystal.


\end{abstract}

 \maketitle

%
\emph{Introduction.} 
Understanding and manipulating nuclear magnetization  in the vicinity of semiconductor-hosted defects is an
issue of  both technological and fundamental importance \cite{Kane1998,Urbaszek2013,Chekhovich2013,Brunner2009}.
It  concerns many different systems, such as semiconductor quantum dots \cite{Urbaszek2013}, nitrogen-vacancy centers in diamond \cite{Togan2011}, silicon-vacancy centers in SiC \cite{Koehl2011}, and other systems, where an electron spin can be used to  transfer angular momentum from light onto nuclei, and nuclei can store information in their spin degree of freedom.
This is possible because nuclear spin system (NSS) is weakly coupled to the lattice \cite{Abragam,OpticalOrientation,DyakonovBook}.
The fact that equilibrium within the NSS is established much faster than the equilibrium with the crystal lattice justifies its thermodynamic description and the concept of nuclear spin temperature $\Theta_N$ \cite{Abragam1958}.
Early studies on  bulk semiconductors demonstrated that combining optical pumping under magnetic fields above local field $B_L$ (given by magnetic interactions within the NSS) with nuclear adiabatic demagnetization to the fields $B<B_L$,  
it is possible to cool the NSS well below lattice temperature \cite{KALEVICH1982,OpticalOrientation,DyakonovBook}.
Various thermodynamic  transitions to spin-ordered states were theoretically  predicted at $\Theta_N<1$~$\mu$K \cite{Merkulov1982,Merkulov98}.
Such degree of control over NSS  would open new possibilities for semiconductor
spintronics, where fluctuations in the nuclear spin system are considered 
as a major and ubiquitous source of decoherence \cite{Imamoglu2003,Reilly2008}.

\begin{figure}
\center{\includegraphics[width=1\linewidth]{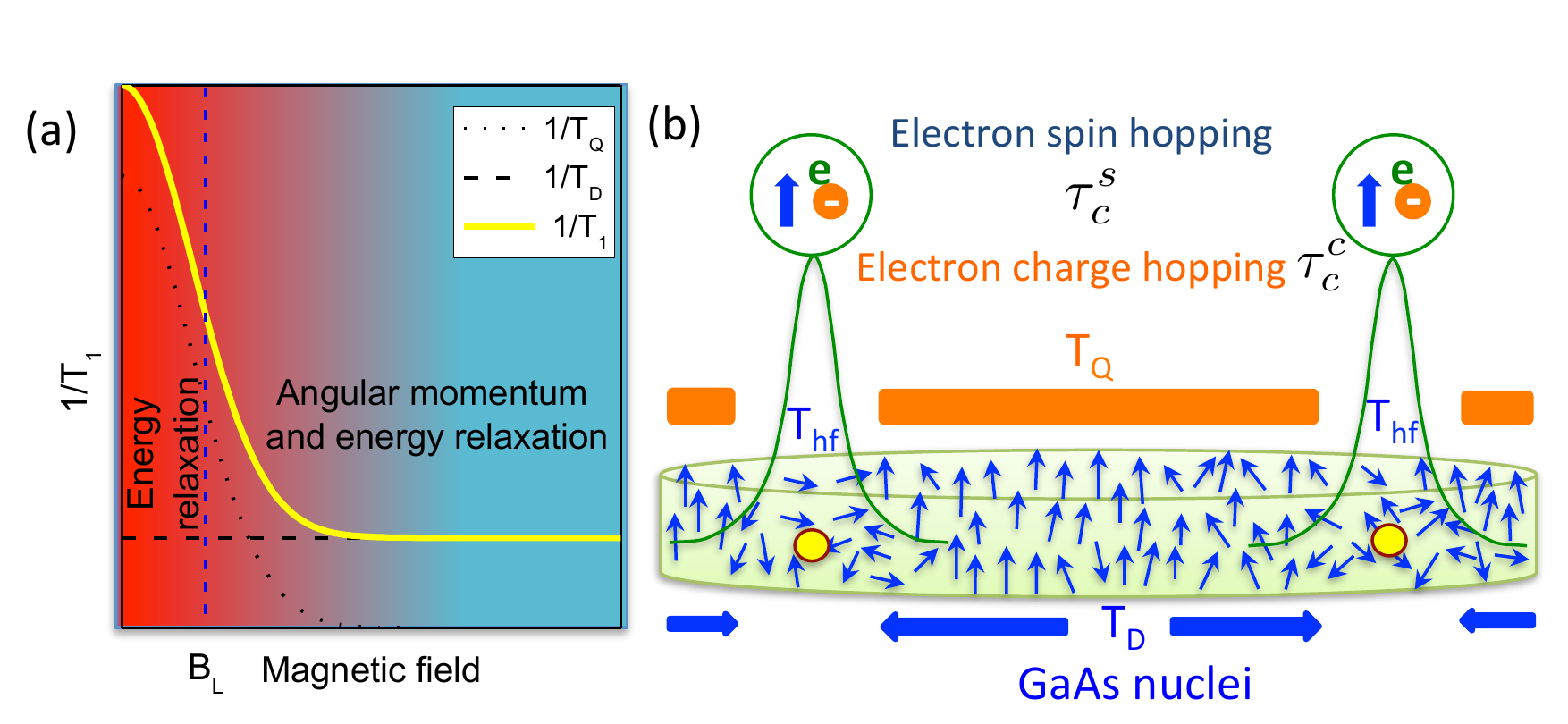} } \caption{
Sketch of the NSS warm-up. (a) Magnetic field dependence of the two contributions to the nuclear spin-lattice relaxation: diffusion-limited hyperfine interaction at rate $1/T_D$ (dashed line) and quadrupole-induced warm up  at rate $1/T_Q$. (b) Relevant processes in nuclei-electrons coupled system: fast nuclear spin warm-up under donor orbits via hyperfine interaction (characterized by time $T_{hf}$), and warm-up of all other nuclei via spin diffusion towards the donor sites and quadrupole interaction.} \label{scheme}
\end{figure}

However, deep cooling of the NSS is challenging, limited by two main issues:
the efficiency of the NSS pumping and the  relaxation during the demagnetisation stage.
It was suggested that light-induced nuclear quadrupolar relaxation can strongly reduce the pumping efficiency \cite{Paget2008}.
Even so, the relaxation of NSS in the absence of optical pumping in a range of magnetic fields down to zero field (i.e. at the conditions met during the demagnetization) has not been addressed so far.
Two different regimes of relaxation should be considered (Fig. \ref{scheme}~(a)).
At strong magnetic fields  the projection of the nuclear angular momentum onto the field is a conserved quantity.
In this regime the energy of the NSS is determined by the Zeeman interaction, so that spin-lattice relaxation involves changing of both energy and angular momentum, mediated by non-spin-conserving interactions.
 At low fields,  $B \le B_L$, the non-equilibrium nuclear angular momentum decays within the spin-spin relaxation time $T_2$ of order of $100$~$\mu$s. 
In this regime the polarization of the NSS is  induced by the external field via its  paramagnetic susceptibility. 
The susceptibility $\chi$ is inversely proportional to $\Theta_N$, which relaxes towards the lattice temperature $T$ with the characteristic spin-lattice relaxation time $T_1>>T_2$. 
 Therefore, the nuclear spin relaxation at low field is in fact the warm-up of the NSS, which is determined by energy transfer between the NSS and the crystal lattice. 
 This warm-up may present a fundamental obstacle on the way towards ultra-low temperatures in NSS. 
 
In this Communication we  investigate the NSS warm-up in bulk n-GaAs using photoluminescence  (PL) spectroscopy at various magnetic fields and temperatures. 
We show that the change of the nuclear spin-lattice relaxation regime from angular momentum relaxation at $B>B_L$ to the NSS warm-up at $B<B_L$ is accompanied by a dramatic increase of the relaxation rate $1/T_1$, Fig.~\ref{scheme}~(a). 
This behavior is completely unexpected\cite{comment1} within the standard model of the  diffusion-limited hyperfine-induced NSS relaxation \cite{comment4,Paget82,Giri2013}.
We suggest that low-field relaxation  is due to  interaction of nuclear quadrupole moments with 
electric field gradients induced by  slow spatio-temporal
fluctuations of localised electron charges,  Fig. \ref{scheme}~(b). 
These fluctuations are characterised by the electron charge correlation time  $\tau_c^c>>T_2$, and 
result from the electron hopping either into conduction band, or between the donor sites, as evidenced by the measurements of resistance as a function of temperature. 
Our theory shows that the energy flux between nuclear spin and electron charge via slowly varying quadrupole interaction $\mathcal{F}_Q$ does not depends on the magnetic field, while NSS heat capacity is strongly field dependent \cite{Goldman}.
This explains the strong field dependence of the NSS warm-up rate. 
The model provides a quantitative understanding of the experimental results and suggest the pathways for the optimisation of the NSS cooling. 
\begin{figure}[t!]
\center{\includegraphics[width=1.0\linewidth]{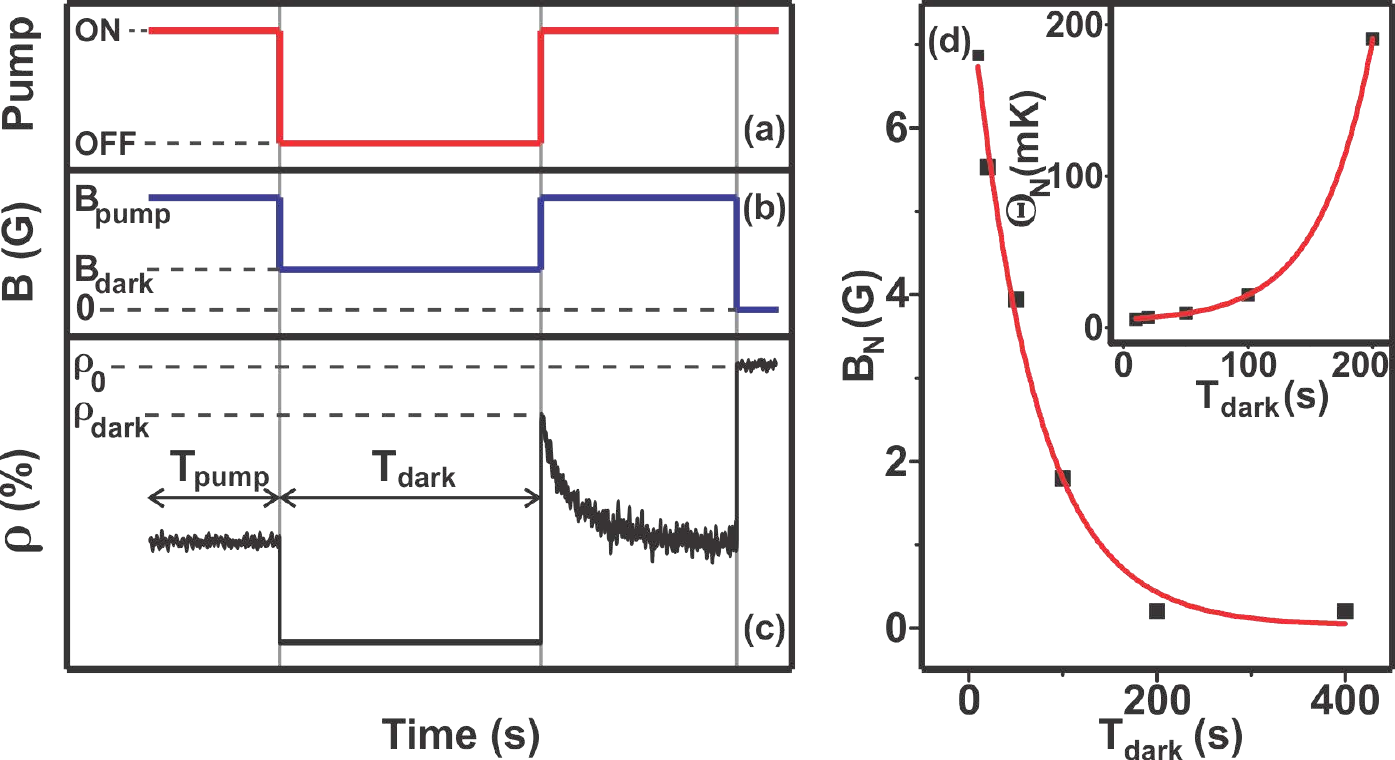} } \caption{
A scheme of the three-stages experimental protocol. Each stage is characterized by (a) the 
presence/absence  of the pump, and  (b) the value of the magnetic field applied during the dark stage. (c) The 
corresponding values of the PL polarisation degree. (d) A set of measurements of $\rho_{dark} \propto B_N$ for a given applied field $B_{dark}=8$~G,  obtained at different values of dark intervals $T_{dark}$ (symbols). The solid line is the fit to the exponential decay, that determines the NSS warm-up time $T_1$. Inset: the same data recalculated in terms of the nuclear spin temperature \label{protocol}}
\end{figure}
\begin{figure}[]
%
%
\center{\includegraphics[width=1\linewidth]{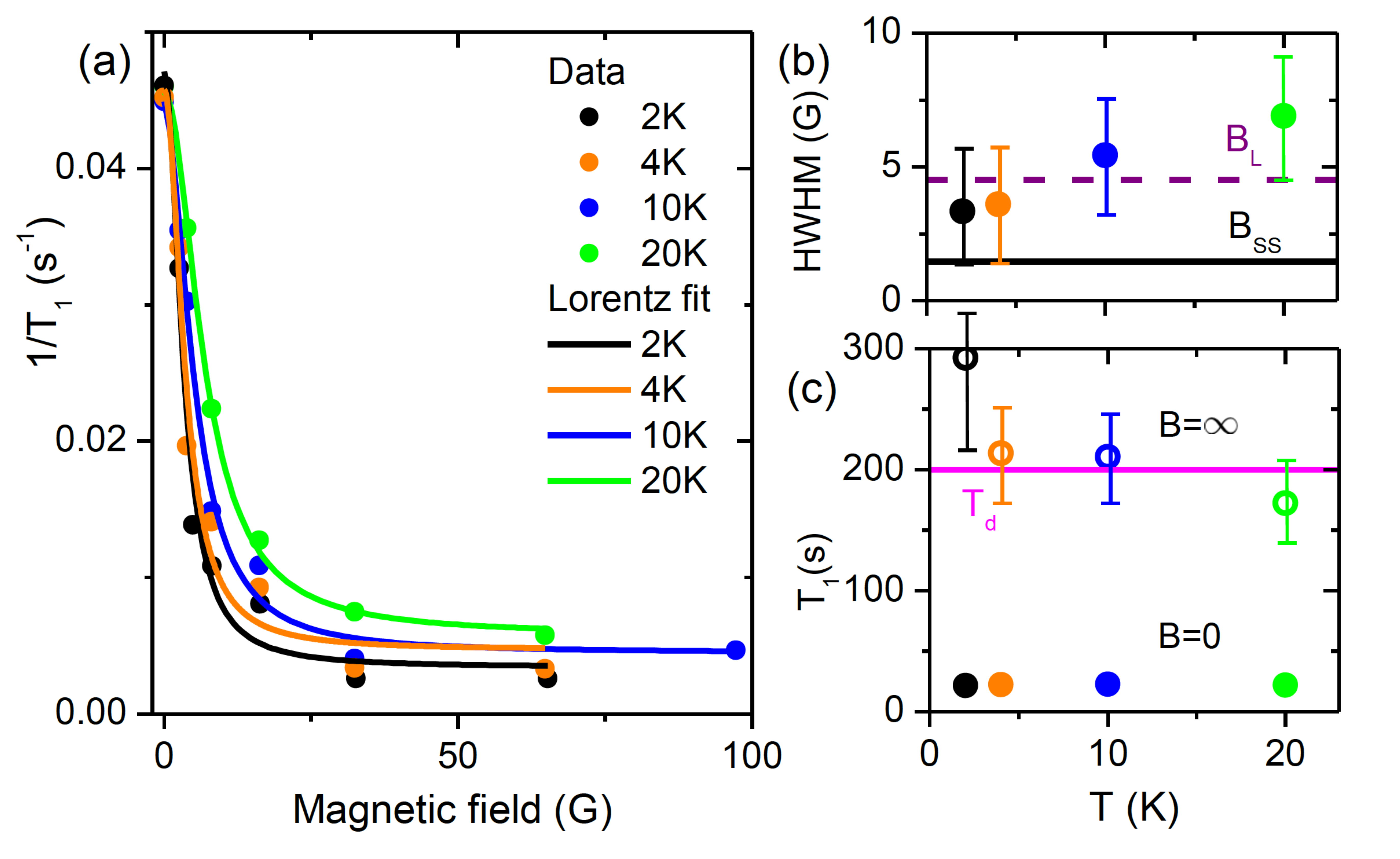} }
\caption{(a) The NSS warm-up rate $1/T_1$ measured as a function of the applied magnetic field (symbols). Solid line are fit to Lorentz function. Values of the HWHM (b), $T_1$ at $B=\infty$ and $B=0$ (c)  extracted from the fitting procedure at  different temperatures. 
 (b) The solid line is the value of local field due to spin-spin interactions, $B_{SS}$  \cite{Paget77}, and the dashed line is the average value of the HWHM extracted from the data, that we interpret as $B_L$.  (c) The solid line is a calculated value of the diffusion limited relaxation time $T_D$.
}
\label{results}
\end{figure}

\emph{Sample and experimental setup.} 
We have studied a $20$~$\mu$m-thick  GaAs sample  grown by liquid-phase epitaxy, with Si donor concentration of $n_d=4\times10^{15}$ cm$^{3}$, already used in a previous work \cite{Dzhioev2002}. The sample was placed in a variable temperature or He bath cryostat, surrounded by an electromagnet. The magnetic field was applied in the oblique but nearly Voigt geometry ($<10 $ degrees), in order to allow for both dynamic nuclear polarisation and detection due to Hanle effect induced by nuclear spin.
The sample was excited by Ti-sapphire laser beam tuned at $E=1.55$~eV, circularly polarised, and  focused on $50$~$\mu$m-diameter spot.
The PL was collected in the reflection geometry, passed through a circular polarization analyzer (consisting of a photoelastic modulator (PEM) and a linear polarizer), spectrally dispersed with a double-granting spectrometer, and detected by an avalanche photodiode, connected to a two-channel photon counter synchronized with the PEM.  %

\emph{Experimental results.} 
The three-stages  measurement  protocol is illustrated in Fig.~\ref{protocol}.
During the pumping stage, both magnetic field $B_{pump}=3.5$~G
and the pumping beam are switched on, providing the cooling of the  NSS.
The duration of this stage is fixed to $T_{pump}=5$~minutes. After $T_{pump}$, the pump beam is switched off,
and the magnetic field is set to the value $B_{dark}$ at which we want to study the warm-up of the NSS.
The second stage of the experiment will be referred to as a dark stage, its duration $T_{dark}$ was varied, in order to access the corresponding variation of the nuclear spin polarization.
Immediately (on the scale of electron spin relaxation time ($T_s\approx 100$~ns \cite{Dzhioev2002}) after the switching off the pump, electron spin polarization returns to its equilibrium state, determined by the sum of the applied magnetic field and the Overhauser field created by the NSS. %
The NSS warm-up is much slower, $T_1>>T_s$.
%
%
In order to measure the evolution of NSS polarisation in the dark, Overhauser field $B_N(T_{dark})$ achieved after $T_{dark}$ is measured during the third stage of the protocol.
The light is switched back on, measuring field  $B_{pump}$ is restored and the circular polarisation degree  of the PL
 is detected as a function of time during $150$~s.
%
%
%
The value of the PL polarisation degree $\rho_{dark}$  in the
beginning of the third stage  is extracted from the exponential fit of this decay, 
for each value of $T_{dark}$ (Fig.~\ref{protocol}~(c)).
The corresponding value of $B_N$, which is proportional to the inverse nuclear spin temperature $\beta=1/k_B \Theta_N$, is  related to  $\rho_{dark}$ via  Hanle formula:
\begin{equation}
\beta(T_{dark}) \propto B_N(T_{dark}) = B_{1/2} \sqrt { \frac {\rho_0-\rho_{dark} }{\rho_{dark}}}-B_{pump},
\end{equation}
where $\rho_0$ is the PL polarisation in the absence of the external field, and $B_{1/2}$ is the half width of the Hanle curve, measured independently \cite{KoturJETP2014} under conditions where nuclear spin polarisation is absent (pump polarisation is modulated at $50$~kHz).
Thus, measuring $\rho_{dark}$  for different waiting times  $T_{dark}$ we can follow nuclear spin warm-up, or, equivalently, the evolution of the Overhauser field $B_N$, Fig.~\ref{protocol}~(d). Fitting the result to the exponential decay
 (growth for $\Theta_N$), we obtain the NSS relaxation time $T_1$ for a given external magnetic field $B_{dark}$ (Fig.~\ref{protocol}~(d))\cite{comment3}.
Similar protocol has been first proposed and realised by Kalevich et al. \cite{KALEVICH1982},  
and then further developed in Ref. \onlinecite{KoturJETP2014}.

\begin{figure}[t]
\center{\includegraphics[width=1\linewidth]{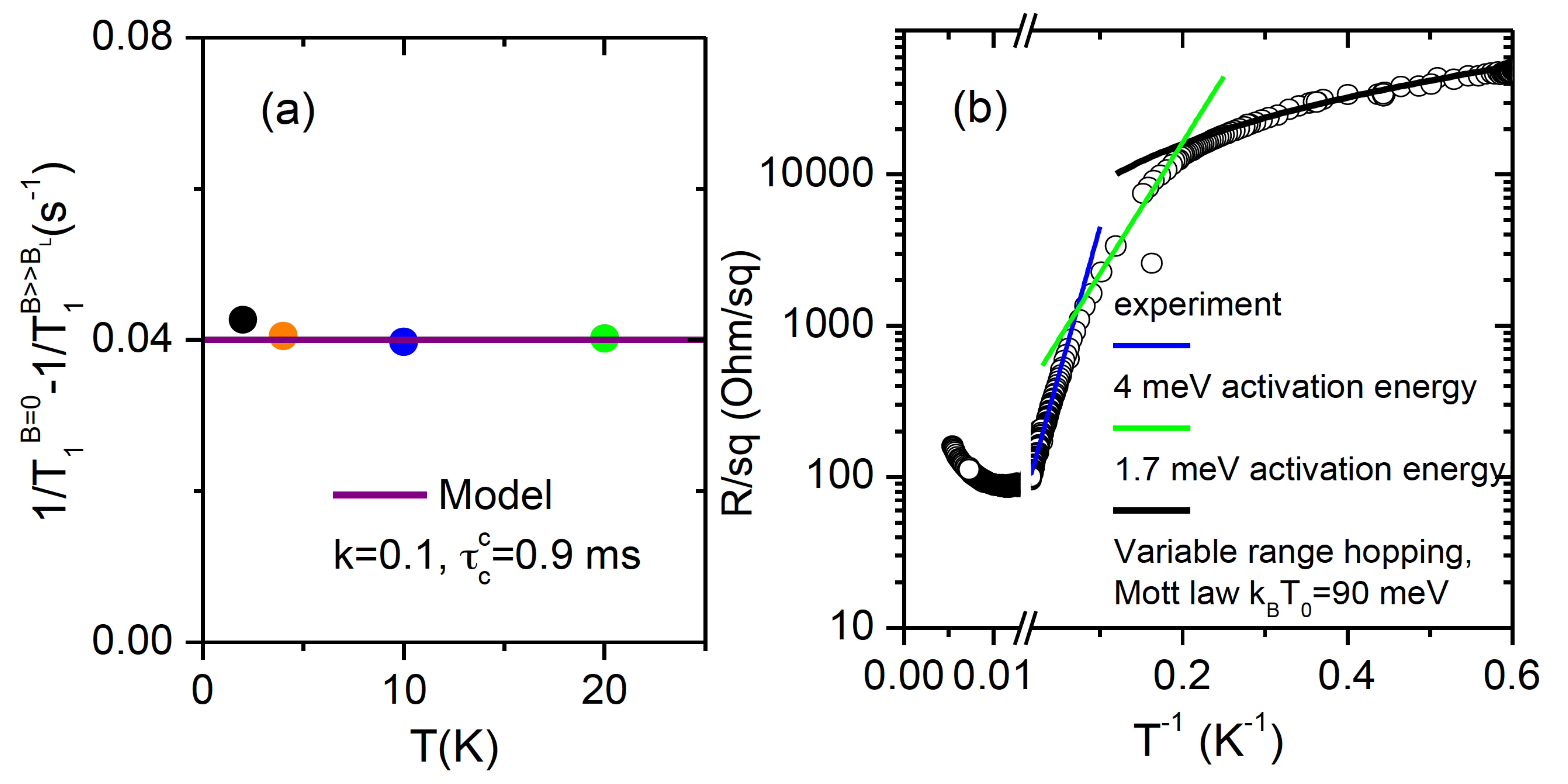} } \caption{
 (a) The enhancement of the NSS warm-up rate measured at different temperatures. The solid line is a fit to Eq. (\ref{T_Q}), with the only fitting parameter  $k/\tau_c^c$. (b) The electrical sheet resistance measured as a function of the  inverse temperature.  Solid lines are fit to the exponential behaviour with two different activation energies, and the Mott law at $T\le5$~K.
}
\label{rho}
\end{figure}

Fig. \ref{results}~(a) summarizes the magnetic field dependence of the NSS warm-up rate $T_1$ measured at different temperatures.
Solid lines are fits to the Lorentz shape with the half width at half maximum (HWHM) reported as color-encoded symbols in Fig. \ref{results}~(b), while the NSS warm-up times  at $B=0$ and $B=\infty$  are reported in  Fig. \ref{results}~(c).
The salient feature of these data is a huge, more than an order of magnitude  enhancement of the NSS warm-up rate with magnetic field changing on the scale of several Gauss.
The HWHM of this dependence,  as well as zero and strong field limits of the  field dependence varie only slightly over the studied temperature
range (see Fig. \ref{results}~(c) and Fig. \ref{rho} (b)).
We present below the model that allows us to describe all these surprising results.

\emph{Theory.}
It is generally admitted that fast  warm-up  of the NSS under the donor orbit via hyperfine interaction, followed by the spin diffusion towards donor sites is the main mechanism of the bulk NSS warm-up  \cite{DeGenes,Paget82,Giri2013}.
Here we deal with another contribution to the bulk NSS warm-up, quadrupole interaction of nuclear spin with fluctuating electric fields.
It results from the energy flux towards NSS via quadrupole interaction ${\mathcal{F}_Q}$:
\begin{equation}
 \frac{1}{T_Q}=\frac{\mathcal{F}_Q}{\beta} \left(C_N\right)^{-1}
 \end{equation}
where $C_N$ is the heat capacity of the nuclear spin system \cite{OpticalOrientation}.
%
%
%
%
Consider a small volume $V$, within which the fluctuating electric field $E_f$ can be assumed spatially homogeneous.
For GaAs and other piezoelectric semiconductors, quadrupole Hamiltonian can be written in the following form:
\begin{equation}
\hat{H}_Q=-V(\vec{E}_f \vec{P}_Q),
\end{equation}
where $\vec{P}_Q$ is the
electric polarization related to the quadrupole magnetic moment. The components 
 of   $\vec{P}_Q$ are given by
\begin{equation}
P_Q^i=\frac{1}{V}\sum_{j k} \beta_{Q}\nu_{j k,i}\sum_n(Q_{j k})_n
\label{PQ}
\end{equation}
Here $\beta_{Q}$ is the experimentally determined and isotope-dependent constant, $eQ$ is  the nuclear quadrupole moment, also isotope-dependent, $e$ is the electron charge, $\nu_{j k,i}=1$ if $i\ne j \ne k \ne i$, and zero otherwise.
Therefore any electric field fluctuating at frequency $\omega$ induces an energy flux $\mathcal{F}_Q(\omega)$ towards nuclear spin system:
\begin{equation}
\mathcal{F}_Q(\omega)=\frac{\omega}{2}\alpha''(\omega)E_f^2(\omega)
\label{Fomega}
\end{equation}
where $\alpha''(\omega)$ is the imaginary part of the generalized susceptibility.
 According to the fluctuation-dissipation theorem (FDT),  $\alpha''(\omega)$   can be expressed through the power spectrum of the fluctuating part of  $\vec{P_Q}$
\begin{equation}
\alpha''(\omega)=V\frac{\omega \beta}{2} < \delta P_Q^2(\omega) >
\label{FDT}
\end{equation}
To calculate the total quadrupole energy flux  $\mathcal{F}_Q=\int_{-\infty}^{\infty} \mathcal{F}_Q(\omega) d\omega$ one needs to account for both electric field and quadrupole polarization fluctuation spectra.
In the absence of illumination, electric field fluctuates due to charge fluctuations induced by thermal activation of donor-bound electrons to the conduction band, or, at lower temperatures, due to hopping to empty donors.
Let us consider the simplest case, when the charge fluctuations are characterized by a single correlation time  $\tau_c^c$. It has a meaning of an average time needed to  fill an empty donor site  by phonon-assisted electron transition from the conduction band, or from a neighbouring donor.
In this case the 
time correlation function can be written as
\begin{equation}
<E_f(0)E_f(t)>=\mathcal{L} E_a^2 e^{-t/\tau_c^c}
\label{correlator}
\end{equation}
Here $E_a$ is the electric field at Bohr radius distance ($a_B$) from the charged donor positon. The dimensionless coefficient  $\mathcal{L}$ accounts for the averaging of the electric fields from the donor-bound electrons $\mathcal{L}=2.5  a_B^3 n_d^+$, where $n_d^+$  is the density of the charged donors \cite{comment2}. 
%
%
Eq. (\ref{correlator}) is valid at time delays larger than the correlation time  $\tau_c^{ph}$  of the fluctuating phonon field, which is usually much shorter than  $\tau_c^c$ (typically  $\tau_c^{ph}$ is in the picosecond range).
Performing the Fourier-transform of Eq.(\ref{correlator}), we obtain the following expression for the power spectrum of the fluctuating electric field:
\begin{equation}
<E_f^2(\omega)>=\mathcal{L}  E_a^2 \frac{2 \tau_c^c}{1+(\omega \tau_c^c)^2},
\label{correlatorFFT}
\end{equation}
valid up to frequencies of the order of order of $2\pi/\tau_c^{ph}$.
Therefore, the quadrupole energy flux  can be expressed as
\begin{equation}
\mathcal{F}_Q=\frac{V \beta}{2 \tau_c^c}\mathcal{L}  E_a^2\int_{-\infty}^{\infty}
 \frac{ (\omega \tau_c^c)^2}{1+(\omega \tau_c^c)^2} <\delta P_Q^2(\omega)> d\omega
\label{flux2}
\end{equation}
Although the exact form of $\delta P_Q^2(\omega)$ is unknown we can take into account that it spreads  mainly in the frequency range 
$\mid  \omega \mid < \gamma_N \sqrt{(B_L^2+B^2)}$, where $\gamma_N$ is nuclear gyromagnetic ratio.
Taking into account that $\tau_c^c >> T_2  \sim 1/( \gamma_N B_L)$ we  can replace the fraction under the
integral by unity, so that
\begin{equation}
\mathcal{F}_Q=\frac{V \beta}{2  \tau_c^c} \mathcal{L}  E_a^2 2 \pi < \delta P_Q^2>
\label{flux3}
\end{equation}
Here $<\delta P_Q^2>$ is the total squared fluctuation of quadrupole polarization. It can be calculated
from quantum mechanical averaging over all nuclear spin states.

Nuclear heat capacity $C_N$ can be written as \cite{OpticalOrientation,Goldman, Wolf}
\begin{equation}C_N=\frac{N}{V}(C_Z+C_{SS}+C_{Q})
\label{HearCapacity}
\end{equation}
where the three terms stand for the heat capacity associated with Zeeman, dipole-dipole, and quadrupole interaction, respectively, $N$ is the total number of nuclei in the volume $V$.
The Zeeman part of the heat capacity is given by 
\begin{equation}
C_Z=I(I+1)(\gamma_N \hbar)^2B^2/3
\label{CZ}
\end{equation}
while  spin-spin, and quadrupole parts are usually expressed in terms of the corresponding effective fields $B_{SS}$ and $B_Q$,
so that $B_L^2=B_{SS}^2+B_Q^2$.
Thus, we obtain the following formula  for the nuclear spin warm-up rate
\begin{equation}
1/T_Q=\frac{4 \pi \mathcal{L}   (eQ \beta_Q E_a)^2}{5 (\hbar \gamma_N)^2 (B^2+B_{L}^2)\tau_c^c}
\frac{4I(I+1)-3}{(8I(2I-1))^2}
\label{T_Q}
\end{equation}
One can see that this rate vanishes at  strong magnetic field but can be  important at low magnetic field. This results from the fact that nuclear heat capacity is strongly field-dependent, while the the quadrupole energy flux is not.
Thus, in the regime where hyperfine relaxation is limited by diffusion, the total warm-up rate is given by
$1/T_1=1/T_Q+1/T_D$, being determined by quadrupole contribution at $B<B_L$ and spin diffusion at strong field, as illustrated by Fig. \ref{scheme}~(a).

\emph{Discussion}
Let us now analyse our experimental data  in the framework of the model.
Hyperfine relaxation rate $T_{hf}$ can be calculated as  $T_{hf}=1/(\omega_{hf} ^2\tau_c^ s)$, where
$\omega_{hf}$ is the nuclear spin precession frequency in the Knight field created by  electron spin localized on the donor site, and 
$\tau_c^s$ is electron spin correlation time \cite{OpticalOrientation}.
In this sample it amounts to  $\omega_{hf}=10^6$~rad/s, $\tau_{c}^s=20$~ps \cite{Dzhioev2002}, $T_{hf}=0.05$~s, so that hyperfine relaxation of bulk nuclei is limited by the spin diffusion rate
$1/T_D=4 \pi D n_d a_B$ \cite{DeGenes}, where $D=10^{-13}$~cm$^2$/s is nuclear spin diffusion coefficient \cite{Paget82}. With $a_B=10$~nm we get $T_D=200$~s, quite close to the experimentally obtained values at strong fields
$B>>B_L$, solid line in Fig. \ref{results}(c).
Spin diffusion rate does not depend neither on temperature nor on magnetic field, so that all the observed variations should be related to quadrupole mechanism.
%

%
%
To compare our model with the experimental results we use the values of $Q \beta_Q=6\times10^{-15}$~cm and $\gamma_N=1.5$~kHz/G averaged
over all three GaAs isotopes, taking into account the abundance of each isotope \cite{Harris2002}. 
Using $E_a=e/a_B^2=12$~kV/cm, $I=3/2$ and $a_B=10$~nm, we end up with three free parameters in Eq.(\ref{T_Q}), $n_d^+$, $B_L$ and $\tau_c^c$.
We assume that our sample has a considerable concentration of acceptors, so that at low temperature the concentration of charged donors $n_d^+\approx k n_d$, where $k$ is the compensation degree. 
It is easy to show that within this model, the HWHM of the $1/T_Q$ field dependence is exactly the value of $B_L$,
while the height of Lorenzian is $1/T_1^{(B=0)}-1/T_1^{(B=\infty)}$.
From this analysis we extract the enhancement of  the NSS relaxation rate shown in Fig.~\ref{rho}~(a) and $B_L=4.5\pm 2$~G, shown by dashed line in  Fig. \ref{results}~(b).
The spin-spin part $B_{SS}=1.5$~G of the local field was calculated and deduced from experiments, its value is shown in Fig. \ref{results}~(b) by solid line \cite{Paget77}.
The missing part of the local field ($B_Q \sim 4$~G) could be attributed to the quadrupole interactions due to uncontrolled strains or to some spin-spin interactions not accounted for in Ref. \onlinecite{Paget77}, such as Dzyaloshinskii-Morya indirect exchange interaction \cite{DM}.
%
%
Assuming the experimentally determined  $B_L=4.5\pm 2$~G, we fit the field-induced enhancement of the warm-up rate by Eq. (\ref{T_Q}). The result is shown as solid line in  Fig.~\ref{rho}~(a), with the fitting parameter 
$k/\tau_c^c=110$~$s^{-1}$.
This means, that for a reasonable compensation degree $k=0.1$, the charge correlation time 
$\tau_c^c=0.9$~ms.
Therefore, our initial assumption $\tau_c^c >T_2$ is \emph{a posteriori} confirmed.
Thus,  the quadrupole-induced relaxation model describes self-consistently the dramatic increase of the NSS warm-up rate at low magnetic fields.
%
%
Expression Eq. (\ref{T_Q}) also suggests that the efficiency of this mechanism can be reduced
in weakly compensated samples.

Since charge correlation time could be related to the resistivity, we also performed sheet resistance measurements in Van der Pauw configuration. 
The result is shown in Fig. \ref{rho}~(b), as a function of the inverse temperature. One can clearly see a non-monotonous behaviour, resulting from combination of the polar optical phonon (above $80$~K) and impurity scattering at lower temperatures \cite{Ehrenreich1960,Wolfe1970}.
In the latter case, we identify  three regimes. 
The resistance between $50$~K and $20$~K  is governed by the thermal activation of bound electrons into the conduction band, the corresponding activation energy $\approx 4$~meV, is of order of the donor binding energy.
From $20$~K to $5$~K electron hopping between
donors takes over, with smaller value of the activation
energy $1.7$~meV.
%
In this regime $\tau_c^c$ should be determined by the hopping between donors. 
Indeed, an estimation of the hopping transition times for typical donor pairs  gives $\tau_c^c=1$~ms at liquid helium temperatures \cite{KavokinSST2008}. 
Below $5$~K, variable range hoping between donors takes over,  giving rise to the Mott law behaviour\cite{ShklovskiiEfros}.
The resulting distribution of the hopping times within the impurity band should be exponentially broad; for this reason, charge fluctuations are expected to have $1/f$ noise spectrum  \cite{Deville2006,Burin2006}.
Thus, the simplified theory operating with a single charge correlation time may be not straightforwardly applicable at the lowest temperatures.
%
%
%
Independent studies of the noise spectra and the generalization  of the model to the case of a distribution of the charge correlation times should allow deeper insight  in the mechanisms of the NSS warm-up rate enhancement at low magnetic field.
%

 %

%
%
%
%

\emph{Conclusions}. We have shown that the warm-up of the NSS  in bulk n-GaAs 
at low magnetic field is governed by previously overlooked mechanism. This mechanism is mediated by the interaction of the quadrupole moment of the nuclei with slowly fluctuating electric fields, due to hopping of the electron charge, either into conduction band, or across the impurity band.
Our analysis suggests that a  possible way to suppress the enhancement of the NSS warm-up at low field, could be found in a careful control and reduction of  the compensation degree.

\emph{Acknowledgements}.  We are grateful to V.~I.~Kozub and D.~Scalbert for
valuable discussions. This work was supported 
%
by the Ministry of Education and Science of the Russian Federation (contract
No. 14.Z50.31.0021 with the Ioffe  Institute, Russian Academy of Sciences,
and leading researcher M. Bayer),
 and by the joint grant  of the Russian Foundation for Basic Research (RFBR) No. 15-52-12017 NNIO-a and
 National Center for Scientific Research (CNRS) PRC SPINCOOL No 148362.




\begin{thebibliography}{56}
\bibitem{comment1}
Decoupling of As and Ga isotopes in strong magnetic fields blocks nuclear
spin diffusion via the first coordination sphere, which may result in approximately
two-fold decrease of the diffusion constant. However, this effect is too small and
occurs in too large magnetic fields (above $20$~G) to explain our observations.
\bibitem{comment4}
This theory states that that nuclear spin relaxation via hyperfine interaction with the localised
electrons is only efficient under the donor orbit, while the warm-up of all other nuclei is controlled
by the spin diffusion towards the donor sites.
\bibitem{comment3}
Since relaxation within the donor Bohr radius is very fast, we
measure $T_1$ of the reservoir of bulk nuclei.
\bibitem{comment2}
Averaging over positions of the probe crystal volume $V$ is equivalent to
the integration of the energy flux over the crystal volume assuming random
spatial distribution of donors over the crystal volume.
\end{thebibliography}
\end{document}